\documentclass[aps,pra,reprint,showpacs,superscriptaddress,onecolumn]{revtex4-2}
\usepackage{braket}
\usepackage{upgreek}
\usepackage{array}
\usepackage{amssymb}
\usepackage{amsmath}
\usepackage{amsfonts}
\usepackage{bbm}
\usepackage{graphicx}%
\usepackage{float} %
\usepackage{subfigure}%
\usepackage{mathrsfs}
\usepackage{graphics,graphicx,epsfig,bm,amsmath,amsthm,amssymb}
\usepackage{bm}
\usepackage{bbm}
\usepackage{longtable}
\usepackage{multirow}
\usepackage{array}
\usepackage{color}
\usepackage{diagbox}
\usepackage[normalem]{ulem}
\usepackage{float}
\usepackage[colorlinks=true,
linkcolor=blue,citecolor=blue,
pdfauthor={ },
pdftitle={ },
pdfsubject={ },
pdfkeywords={ }]{hyperref}

\begin{document}

\title{Active Control of Topological Exceptional Points in Non-Hermitian Metasurfaces}
\author{Parul Sharma}
\thanks{These authors contributed equally to this work.}
\affiliation{Laboratory of Optics of Quantum Materials, Department of Physics, Indian Institute of Technology Bombay (IITB), Mumbai -- 400076, India}
\author{Sobhan Subhra Mishra}
\thanks{These authors contributed equally to this work.}
\affiliation{Division of Physics and Applied Physics, School of Physical and Mathematical Sciences, Nanyang Technological University, Singapore -- 637371}
\author{Yash Gupta}
\affiliation{Laboratory of Optics of Quantum Materials, Department of Physics, Indian Institute of Technology Bombay (IITB), Mumbai -- 400076, India}
\author{Brijesh Kumar}
\affiliation{Laboratory of Optics of Quantum Materials, Department of Physics, Indian Institute of Technology Bombay (IITB), Mumbai -- 400076, India}
\author{Ranjan Singh}
\email{rsingh3@nd.edu}
\affiliation{Department of Electrical Engineering, University of Notre Dame, Notre Dame, USA}
\author{Abhishek Kumar}
\email{abhishekkumar@jncasr.ac.in}
\affiliation{School of Advanced Materials (SAMaT, Jawaharlal Nehru Centre for Advanced Scientific Research (JNCASR), Bangalore, 560064, India
}
\affiliation{Chemistry and Physics of Materials Unit (CPMU), Jawaharlal Nehru Centre for Advanced Scientific Research (JNCASR), Bangalore, 560064, India}

\author{Anshuman Kumar}
\email{anshuman.kumar@iitb.ac.in}
\affiliation{Laboratory of Optics of Quantum Materials, Department of Physics, Indian Institute of Technology Bombay (IITB), Mumbai -- 400076, India}
\affiliation{Centre of Excellence in Quantum Information, Computation, Science and Technology (QuICST), Indian Institute of Technology Bombay, Mumbai -- 400076, India}

\begin{abstract}  
Active control and ultrafast switching of non-Hermitian photonic systems are essential for next-generation reconfigurable optical technologies. 
Here, we demonstrate {dynamic} temporal manipulation of EPs in the terahertz (THz) regime using optically excited germanium (Ge) as an active medium. By exploiting pump–probe delay as a continuous tuning parameter, we achieve sub-picosecond eigenmode switching ($\sim0.5$ ps) and realize a complete time-resolved EP encirclement within $\sim2$ ps, enabling direct observation of topological phase accumulation. At EP, the metasurface exhibits highly asymmetric transmission for circularly polarized light, characteristic of chiral mode response. Furthermore, we observe ultrafast eigenmode switching and topological phase evolution within $\sim1$ ps, achieving $>99\%$ cross-polarization modulation depth. The measured results show strong agreement with theoretical modelling, with a high Petermann factor $\approx10^3$, confirming the effectiveness of the design. Our work establishes pump–probe delay {time} as a {dynamical} control parameter for EP topology, introducing a new regime of ultrafast non-Hermitian photonics for high-speed switching, enhanced sensitivity, and tunable polarization control in the THz domain.

\end{abstract}
\maketitle
\section{Introduction}
Non-Hermitian physics has emerged as a powerful paradigm for engineering exotic spectral and dynamical behaviour in photonic systems, including square root topological transitions~\cite{Lin2021}, non-Hermitian Skin effect~\cite{Zhang2022,Lin2023, Ma2024}, unidirectional invisibility~\cite{Lin2011}, loss-induced lasing and suppression~\cite{Peng2014}, optomechanically induced squeezing~\cite{delPino2022}, single photon interferometry~\cite{Wang2021_SPI}, many-body topological excitations~\cite{Hyart2022} and formation of bulk Fermi arcs~\cite{Zhou2018}. Central to these phenomena are exceptional points (EPs), which are singularities where both eigenvalues and eigenvectors coalesce, leading to branch-point topology in parameter space~\cite{Heiss2012,Miri2019}. Over the past decade, EPs have enabled a broad range of photonic phenomena, including loss-induced transparency~\cite{L2018,Wang2020}, non-reciprocal mode conversion~\cite{Doppler2016, Lee2025}, enhanced sensitivity~\cite{Chen2017, Sharma2024}, topological energy transfer~\cite{Xu2016,Wang2021}, and chiral state evolution~\cite{Kang2016,Sweeney2019}. Yet, despite intense theoretical and experimental activity, EP-enabled dynamics have mostly been studied with static or quasi-static parameter control~\cite{Doppler2016,Song2021,Zhang2023}. A major frontier gaining recent interest is ultrafast parameteric encirclement of an EP, in which a system is driven around an EP on sub-picosecond timescales. 
This regime enables topologically structured mode switching~\cite{Ding2022}, high-efficiency asymmetric mode conversion, and opens new opportunities for ultrasensitive biomolecule sensing, chiral enantiomer discrimination, and the development of compact photonic devices.

Terahertz (THz) regime offers a particularly promising platform for exploring dynamically tunable non-Hermitian physics\cite{Li2023}. This spectral range is especially advantageous due to the relative ease of fabricating high-precision micron-scale metasurfaces, as well as its suitability for pump–probe experiments, where the intrinsically sub-picosecond resolution of THz pulses enables direct probing of ultrafast dynamics on comparable timescales. THz metasurfaces can be engineered to support sharply defined resonances with strong field confinement and enhanced overlap with active materials, while simultaneously allowing precise control over radiative and absorptive loss channels\cite{Yu2024}. Their electromagnetic response is highly sensitive to conductivity variations in photoexcited semiconductors, enabling active tuning with relatively modest optical pump fluences~\cite{Lim2018}. Despite these advantages, most previous demonstrations of EP control in the THz domain have been limited by slow carrier relaxation or thermally driven processes, restricting parameter modulation to nanosecond or longer timescales~\cite{Doppler2016,Liu2021,Shu2022,Zhang2023}. Such constraints prevent access to the ultrafast regimes required for real-time EP manipulation and {dynamical} {parametric} encirclement~\cite{Milburn2015,Choi2020}. Furthermore, implementing simultaneous and independent control over multiple system , such as losses and resonant frequencies, is essential for traversing closed loops around EP in parameter space, and remains a significant experimental challenge in THz platforms~\cite{Shu2022}.

In this work, we address these limitations by integrating a 100 nm germanium (Ge) layer into a carefully engineered THz metasurface, exploiting the intrinsically short carrier lifetime of Ge to achieve picosecond-scale modulation of optical loss under optical pump with THz probe excitation.
Compared to previous approaches~\cite{Weibao2023, Zhongyi2024}, our platform achieves significantly faster {parametric} EP encirclement {dynamics}, enabling ultrafast eigenmode switching and topological phase evolution on the $\sim1$ ps timescale, while maintaining a cross-polarization modulation depth exceeding 99$\%$ during the encirclement path. Furthermore, the proposed design simplifies fabrication, as the Ge layer is uniformly deposited across the metasurface without requiring additional lithographic steps. We have also shown how the non-ideal behavior of THz optics, such as THz polarizers can result in a shift of EP position in parameter space.
Photoexcitation of the Ge layer introduces a rapidly tunable non-Hermitian loss channel whose temporal evolution closely follows the pump–probe delay, with recombination occurring between the photoexcited electron-hole pairs within only a few picoseconds. This ultrafast response enables the realization of time-dependent trajectories in parameter space, where both the dissipative and dispersive properties of the metasurface evolve dynamically.

By leveraging this ultrafast tunability, we achieved a high modulation depth (MD) in the cross-polarization transmission channel and demonstrated dynamic navigation to and from EP conditions, directly observing EP-induced chiral mode switching behaviour as highlighted in \ref{fig:schematic} (a). We showed two sets of parameters for fine-tuning the system around EP, the first set being the THz probe frequency and pump fluence, which enables active control of the system by changing laser power, and the second set being the pump-probe delay, which enables us to measure and control the ultrafast polarization dynamics of the transmitted beam.

When the system parameters are tuned around EP, the metasurface exhibits strongly asymmetric transmission for incident circularly polarized beam, reflecting the coalescence of eigenvalues and eigenstates and the emergence of a chiral response. Beyond static tuning, the rapidly decaying carrier population in Ge provides a built-in temporal dimension that allows EP encirclement to be realized on picosecond timescales. The resulting state evolution reveals ultrafast eigenmode switching driven by non-Hermitian topology, providing direct experimental evidence of EP dynamics.
A key feature of this platform is the exceptionally small eigenvalue and eigenstate splitting at the coalescence point, yielding a large Petermann factor around $10^3$, indicating pronounced eigenstate non-orthogonality. This evidences a near-complete modal coalescence enabled by finely balanced coupling and loss, establishing an optically tunable THz metasurface as a compelling platform for realizing and exploiting strong non-Hermitian degeneracies. This near degeneracy significantly enhances the system’s sensitivity to external perturbations and highlights the potential of EP-enabled functionalities for ultrafast switching and sensing applications.

Overall, our results establish a new regime of non-Hermitian photonics in which EPs can be actively manipulated on sub-picosecond timescales in the THz domain. The demonstrated ability to dynamically control loss, induce chiral mode evolution, and achieve ultrafast EP-driven switching paves the way for reconfigurable THz devices, including high-speed modulators, polarization-selective components, and time-varying metasurfaces that harness the unique topology of exceptional points.

\section{Results}
\subsection{Theoretical Modelling for Dynamic EP control}

\begin{figure*}[htb]
    \centering
    \includegraphics[width=0.8\textwidth]{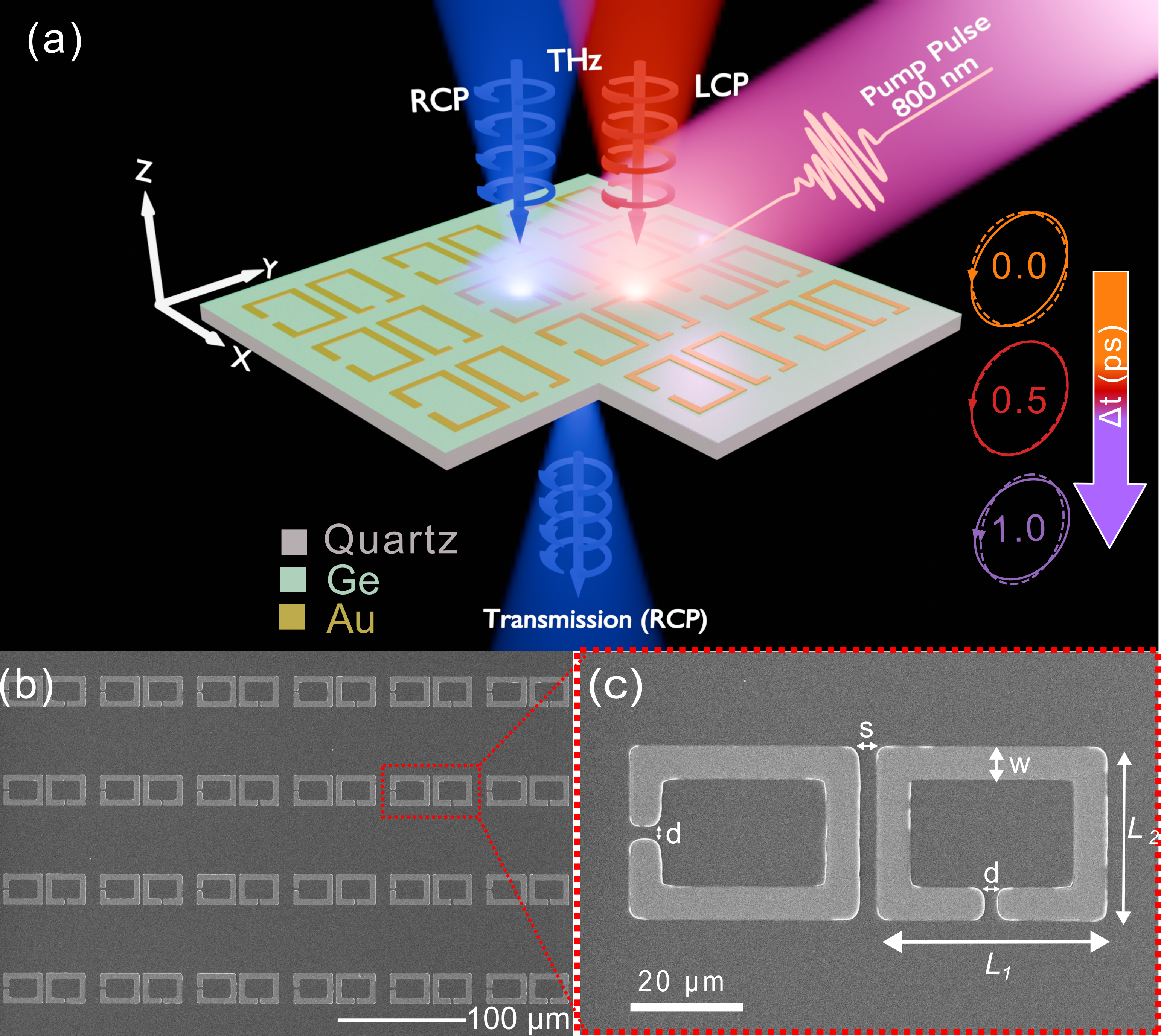}
   \caption{\textbf{Metasurface exhibiting asymmetric circular polarization conversion and ultrafast dynamics. (a)} Schematic representation of the metasurface, illustrating the temporal evolution of the polarization state via polarization ellipses, highlighting ultrafast switching dynamics on a sub-picosecond timescale. \textbf{(b)} Scanning electron microscopy (SEM) image of the fabricated metasurface. \textbf{(c)} Magnified view of an individual square unit cell with periodicity $P_x = P_y = 100~\mu\text{m}$, $L_1 = 40~\mu\text{m}$, $L_2 = 30~\mu\text{m}$, $w = 5~\mu\text{m}$, $s = 3~\mu\text{m}$, and $d = 2~\mu\text{m}$.}
    \label{fig:schematic}
\end{figure*}

To study the time-dependent parameters, we construct a non-Hermitian Hamiltonian $\mathcal{H}$ in the Cartesian basis. We link the x and y oriented resonator's dipole moment state vector $\bm{p}(t)=\left[p_x(t),p_y(t)\right]^T$, whose dynamics is driven by the incoming driving electric field $\bm{E}_{\rm{in}}(t)=\left[E_x(t),E_y(t)\right]^T$. The system can be modelled by an effective time-dependent Schr\"{o}dinger equation $i\partial_t \bm{p}(t) = \mathcal{H}\bm{p}(t) + g\bm{E}_{\rm{in}}(t)$ using temporal coupled mode theory (TCMT), for $\mathcal{H}$ as
\begin{align}
\mathcal{H}=\begin{pmatrix}
\omega_{x} -i\gamma_x - G_{xx} & -G_{xy}\\
-G_{xy} & \omega_{y} -i\gamma_y - G_{yy}
\end{pmatrix},
\label{eq:Ham_reso}
\end{align}
where $\omega_{x(y)},\gamma_{x(y)}$ are the resonant frequency and loss associated with the x(y) oriented resonators. $G_{ij}$ are the summation of all dipole-dipole coupling strengths for $i$ oriented resonator dipole to $j$ oriented one, $i,j\in\{x,y\}$. In our case, a square unit cell geometry ensures real reciprocal couplings in off-diagonal elements, therefore, $G_{xy}=G_{yx}$ cite{Park2020}. The driving field is coupled to the dipole moment via interaction $g=\rm{Diag}\left(\left[g_x,g_y\right]\right)$ which depends on the split gaps of the resonators {and the length of the resonator along the dipole}.
Figure~\ref{fig:schematic}(a) presents a schematic illustration of the THz metasurface, highlighting the device geometry, the asymmetric state transition, and the ultrafast eigenvalue switching. The bottom-left inset shows an SEM image of the fabricated sample, while the central part depicts the layered structure, with the Ge layer encapsulating the gold metasurface. Notably, this complete encapsulation eliminates second-level fabrication, making the process simpler and more streamlined compared to previous studies~\cite{Lim2018-vb,Weibao2023,Zhongyi2024}. The right inset illustrates the ultrafast parametric switching, where the polarization ellipses corresponding to the two eigenstates evolve and coalesce at EP at a delay time of $\Delta t = 0.5$ ps. Fig.~\ref{fig:schematic}(b) shows a zoomed-in SEM image of the metasurface.
We designed the resonator geometry to be rectangular ($L_1 \neq L_2$) in order to introduce anisotropic losses even under uniform optical pumping. The current distribution along the dipoles differs between the long and short axes because the effective inductance and capacitance are unequal in the two directions, leading to direction-dependent electromagnetic response.
We now express the frequency response for the electric field and dipole oscillating with $\bm{E}(t)=\Tilde{\bm{E}}e^{-i\omega t},\bm{p}(t)=\Tilde{\bm{p}}e^{-i\omega t}$. This allows us to rewrite Eq.~\ref{eq:Ham_reso} as $g\Tilde{\bm{p}}=\alpha(\omega) \Tilde{\bm{E}}_{\rm{in}}$, where $\alpha(\omega)=g\left(\omega I - \mathcal{H}\right)^{-1} g$ is the polarizability matrix, which is directly related to the spectral resolvent of the system Hamiltonian $\mathcal{H}$. Now the transmitted electric field is obtained by superposing the input field with the field radiated by the induced dipoles, normally to the surface, $\Tilde{\bm{E}}_r=-i(\omega\eta_0/2a^2) g\Tilde{\bm{p}}$, thus, $\bm{\tilde{E}}_t=\bm{\tilde{E}}_{\rm{in}} + \bm{\tilde{E}}_r=T(\omega)\bm{\tilde{E}}_{\rm{in}}$, where $T(\omega)$ is the transmission matrix.
\begin{align}
T(\omega)=I-\frac{i\omega\eta_0}{2a^2}g\left(\omega I - \mathcal{H}\right)^{-1} g=I-\frac{i\omega\eta_0}{2a^2}\alpha(\omega)
\label{eq:Etrans}
\end{align}
where $a$ is the periodicity of unit cell, $\eta_0=\sqrt{\mu_0/\epsilon_0}$ is the free space impedance, 
$T(\omega)$ is the transmission matrix carrying various transmission spectrum components $T_{i,j}(\omega)$, which can be calculated in experiments. The conditions for the emergence of EPs can be determined from the transmission eigenvalues, $\lambda_{\pm}(\omega)$ of $T(\omega)$ by requiring the eigenvalue gap, $\Delta\lambda=\lambda_+-\lambda_-$ to vanish. This yields the exact conditions, $\Delta\omega=0$ and $\Delta\gamma=\pm G_{xy}$,under which $\alpha(\omega)$ becomes a defective matrix. Here $\delta_{x(y)}=\omega-\omega_{x(y)}$ are the detunings with respect to probe frequency, while $\Delta\omega=(g_x^2\delta_y -g_y^2\delta_x)/2g_xg_y,~\Delta\gamma=(g_y^2\gamma_x-g_x^2\gamma_y)/2g_xg_y$ are the weighted differences in detuning and loss, respectively. A detailed derivation of eigenvalues, eigenstates and realization of EP condition is presented in SM~\cite{SM}~S5. In our case, the inter-unit cell coupling strengths are weak, hence $G_{xx},G_{yy}\approx 0$.

At EP, the right eigenstates can be obtained as
\begin{align}
    \vert\psi_{\pm}\rangle^{\rm{EP}}=\frac{1}{\sqrt{\vert\Delta\gamma\vert^2 + \vert G_{xy}\vert^2}}\begin{pmatrix}
        \Delta\gamma\\
        iG_{xy}
    \end{pmatrix}=\frac{1}{\sqrt{2}}\begin{pmatrix}
        1\\
        s~i
    \end{pmatrix}
\end{align}
where, $s=\rm{sgn(\Delta\gamma/G_{xy})}$. Thus, at EP, both eigenstates merge into either LCP or RCP state depending on whether the sign $s$ is $-1$ or $+1$. In our experiment, we have $s=-1$ leading to LCP as merged eigenstates at EP.

\begin{figure*}[htb]
    \centering
    \includegraphics[width=\textwidth]{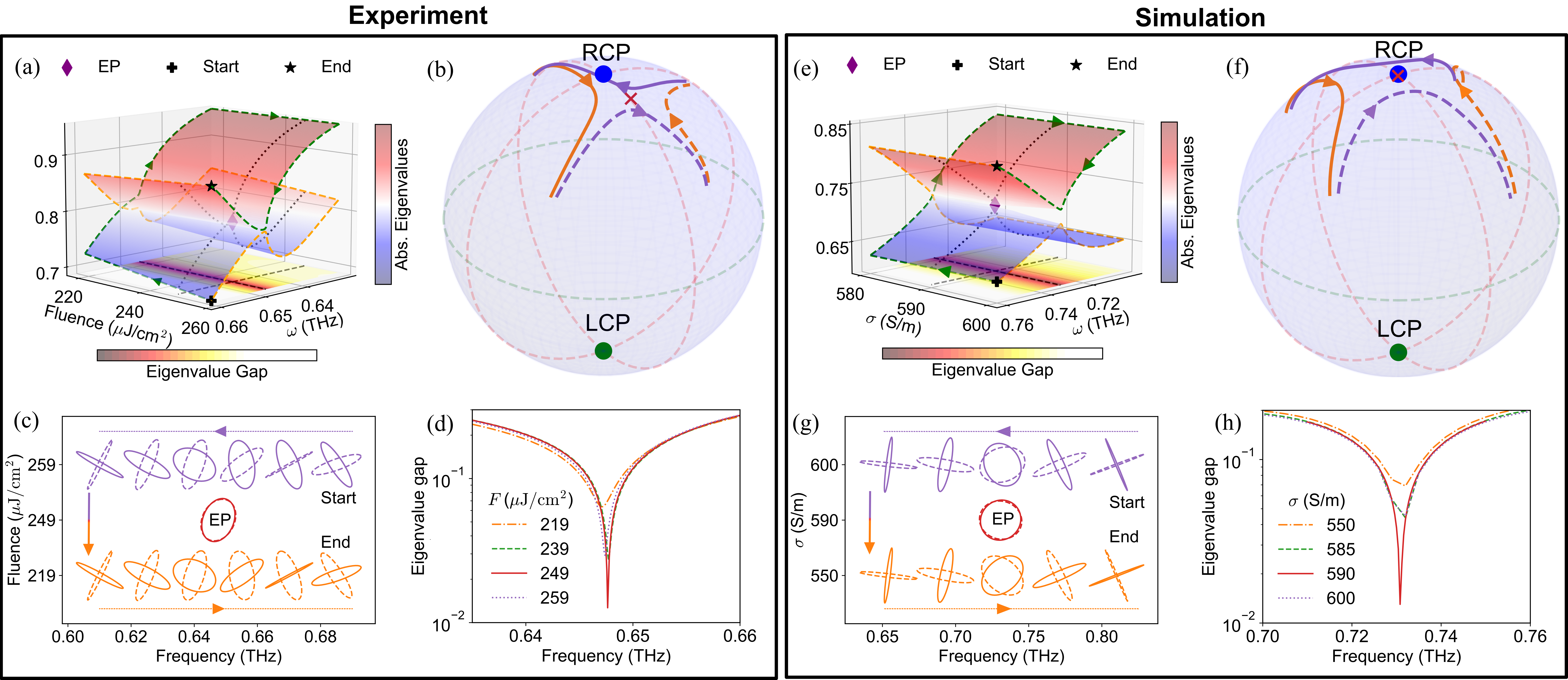}
    \caption{\textbf{Active evolution of eigenstates around EP: (a,e)} Encirclement of EP on the absolute transmission eigensurfaces, realized by continuously varying the parameter pairs $(\omega, F)$ in the experiment and $(\omega,\sigma)$ in simulations, respectively. The closed trajectories in parameter space traverse a non-Hermitian degeneracy associated with the EP. \textbf{(b,f)} Corresponding encirclement paths are mapped onto the Poincaré sphere, revealing the continuous evolution of the polarization eigenstates during the parameter sweep. \textbf{(c,g)} Topological eigenstate swapping induced by a complete loop around the EP, visualized through the evolution of polarization ellipses. The exchange of eigenstates after one encirclement is a hallmark of EP-associated non-Hermitian topology. \textbf{(d,h)} Progressive collapse of the eigenvalue splitting as the system parameters approach the EP, demonstrating the vanishing eigenvalue gap in both experiment and simulation.}
    \label{fig:sims}
\end{figure*}
Our primary objective is to actively tune the system around exceptional point without requiring refabrication or any modification of geometrical parameters, thereby minimizing fabrication-induced errors and improving reproducibility. To achieve this, the thin Ge layer on the metasurface is optically pumped with varying laser powers, thereby modulating the conductivity $\sigma$. As $\sigma$ is swept, the resonant frequencies $\omega_{x(y)}(\sigma)$, decay rates $\gamma_{x(y)}(\sigma)$, and coupling coefficients $g_{x(y)}(\sigma)$ vary approximately linearly with conductivity; the fitting and linear variation results are shown in SM~\cite{SM} S3, and S6. In contrast, the dipole–dipole coupling terms $G_{ij}$ and resonance frequencies $\omega_x(y)$ remain effectively unchanged, as confirmed by curve fitting of both experimental and simulated results using TCMT.
This selective tunability enables controlled adjustment of the weighted differential loss $(\Delta \gamma)$ between the two resonators relative to the coupling strength $G_{xy}$.

\subsection{Active tuning of EP through photoexcitation}
In this work, we demonstrate active {temporal} control and {dynamic} {parametric} encirclement of an exceptional point (EP), leading to eigenstate swapping without altering the metasurface geometry. Encircling an EP requires access to a two-dimensional parameter space. In our approach, the transmission frequency $\omega$ of the output signal naturally provides one effective (synthetic) dimension, as the complete transmission spectrum is measured. This allows EP to be encircled by introducing control over a second independent parameter, namely the optical pump fluence.
To realize this experimentally, to achieve this, we employ an optical-pump terahertz-probe (OPTP) configuration (see supplementary material (SM)~\cite{SM}~S1), which enables modulation of the carrier density of photo-excited electron–hole pairs in the Ge substrate while simultaneously measuring the THz transmission spectrum under varying optical pump conditions. The key parameter controlling the conductivity of the Ge layer is the incident fluence ($F$), delivered through an 8 mm aperture onto the metasurface.
By sweeping the transmission frequency for different values of $F$, we trace a closed rectangular loop in the $(F,\omega)$ parameter space, thereby encircling the EP, as illustrated in Fig.~\ref{fig:sims}(a,e). All the components of experimentally measured and simulated transmission components are shown in SM~\cite{SM} S2. Due to the nontrivial topology of the eigenvalue surface, the endpoint (black star) lies on a different eigenvalue surface after one complete traversal of the loop (green dashed curve), which originates from the initial point marked by a black plus. The colormap on the surface represents the magnitude of the eigenvalues centered at the EP, indicated by a violet diamond. Cross-sections passing through the EP are denoted by black dashed and dotted lines, while the lower colormap shows the variation of the eigenvalue gap across the parameter space.
Figures~\ref{fig:sims}(b,f) illustrate the evolution of the eigenstates on the Poincaré sphere for the two states $\vert\psi_{\pm}\rangle$, represented by solid and dashed trajectories, respectively. The directional arrows indicate the encirclement path, clearly showing that the eigenstates interchange upon completion of the loop. In experiments, the eigenstates deviate slightly from the ideal left circular polarization (LCP) state predicted in simulations due to the presence of an imperfect polarizer (see SM~S4). This leads to a shift in EP position towards lower frequency and conductivity compared to the ideal case.
A more explicit visualization of the eigenstate evolution is provided by the polarization ellipses shown in Fig.~\ref{fig:sims}(c,g). In particular, at optimal EP parameters, both the experimental results (Fig.~\ref{fig:sims}(d)) and simulations (Fig.~\ref{fig:sims}(h)) exhibit a transmission eigenvalue gap ($\Delta\lambda$) of the order of $\approx 10^{-2}$.

\begin{figure*}[t]
    \centering
    \includegraphics[width=\textwidth]{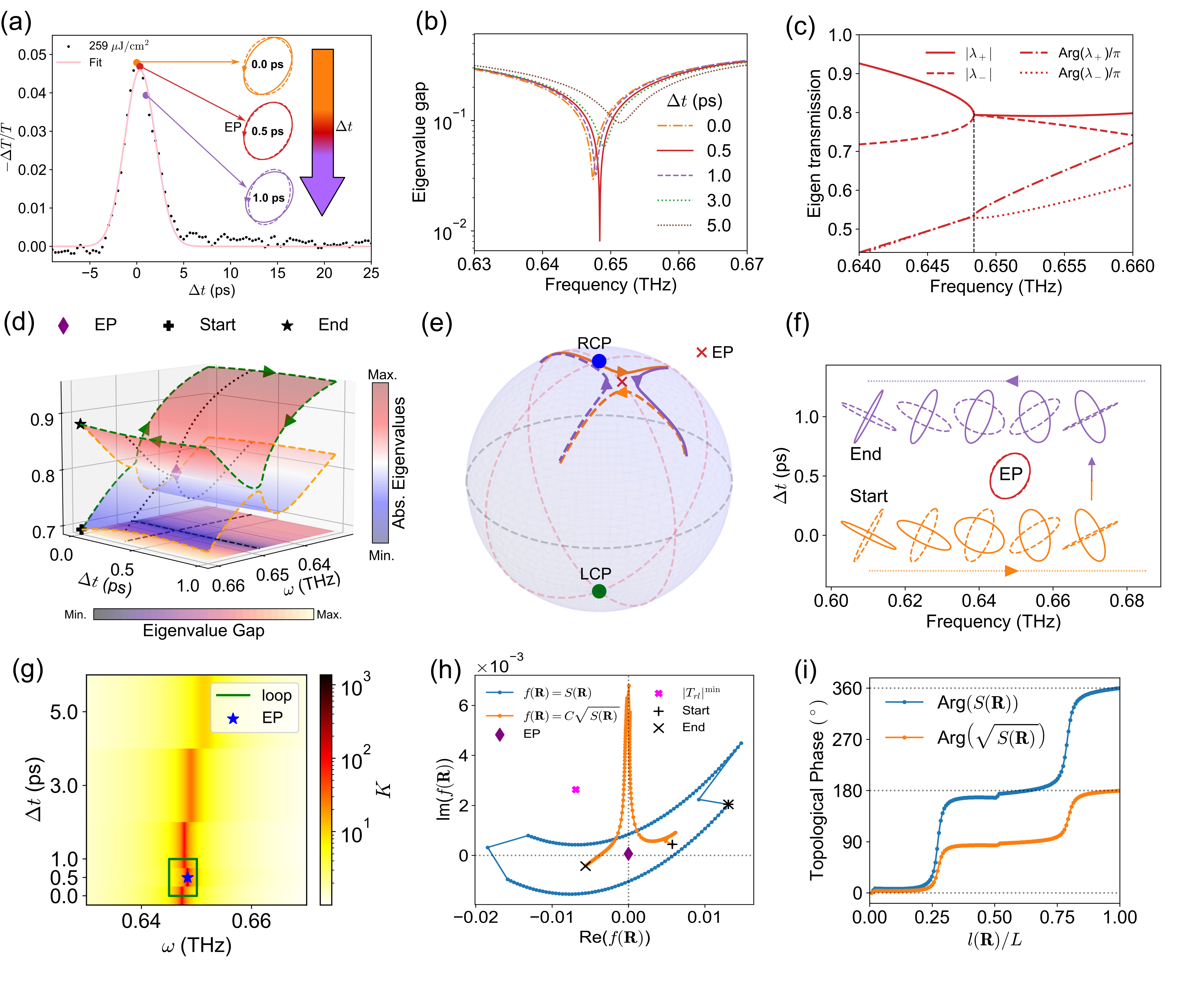}
    \caption{\textbf{Ultrafast mode switching: (a)} Carrier response dynamics of Ge showing switching through EP at a delay time of 0.5 ps. \textbf{(b)} Temporal evolution of eigenvalue gap $\Delta\lambda\rightarrow0$ as the system approaches EP. \textbf{(c)} Eigenvalues $\lambda_{\pm}$ coalesce at EP $\omega_{\rm{EP}}=0.6483$ THz. \textbf{(d-f)} Encirclement around EP by varying probe frequency $\omega$ and pump delay $\Delta t$ for the fixed fluence $259~ \mu$J/cm$^2$. \textbf{(g)} Experimental values of Petermann factor $K$, reaching its maxima $(K\approx1365)$ near EP marked as blue star. \textbf{(h)} Encirclement path (green rectangular loop in (g)) for $S(\mathbf{R})=\Delta\lambda^2(\bm{R})$, where $\Delta\lambda$ is the eigenvalue gap $\sqrt{S(\mathbf{R})}$ in the complex plane spanned by parameter space vector $\bm{R}=(\omega,\Delta t)^T$. The starting point is marked as plus, and the ending point is marked as cross. \textbf{(i)} Accumulated topological phase variation with normalised path length $l(\mathbf{R})/L$ while encircling the loop shown in (g). $C=1/20$ is a scaling factor introduced for better visibility in (h).}
    \label{fig:exp_fast}
\end{figure*}

The precise tunability of the optical fluence enables fine control over the system parameters, allowing access to regions in close proximity to the EP where both the eigenvalues and eigenstates separations become vanishingly small. This high degree of control highlights the sensitivity and robustness of our platform, providing a powerful means to probe EP physics with remarkable proximity and tunability.

\subsection{Sub-Picosecond switching to EP Using Ge carrier dynamics}
Germanium (Ge) is particularly well suited for ultrafast modulation in non-Hermitian photonic systems due to its strong optical absorption and highly tunable carrier dynamics in the THz regime. Upon optical excitation, with an 800 nm ($1.5~eV$) pump whose photon energy exceeds the bandgap of Ge ($0.66~eV$), free carriers are efficiently generated. The optical pump beam has a diameter of 8 mm, while THz probe beam is confined to a 4 mm spot, ensuring that the probed region lies well within the uniformly illuminated area, thereby eliminating spatial inhomogeneity. The subsequent carrier recombination dynamics govern the temporal evolution of this conductivity and, consequently, the material’s electromagnetic response. The modulation in longitudinal transmission arising from these carrier dynamics can be described by the following equation

\begin{align}
    \frac{\vert\Delta T\vert}{T}(t)= e^{-(t-t_0)^2/\sigma_\tau^2}\left[A_0 + A_1 e^{(t-t_0)/\tau_1}\right],
\label{eq:dT/T_fit}
\end{align}
where T is intrinsic THz transmission through the sample without pump along the x-axis, $\Delta T=T_{\rm{on}}-T$ is the change in transmission in the presence of pump pulse, $\sigma_\tau$ is the Gaussian envelope width representing the instrument response function and $\tau_1$ is the effective relaxation time of Ge carriers with $A_0, A_1$ as the amplitudes which can be calculated by curve fitting. We fit this function to the experimentally obtained transmission as shown in Fig.~\ref{fig:exp_fast}(a). Notably, we have used single exponential form in Eq.~\eqref{eq:dT/T_fit}, similar to~\cite {Lim2018}.

This ultrafast carrier relaxation allows the conductivity of Ge to be dynamically modulated as the optical pump pulse exits the sample. Consequently, the system parameters can be switched on timescales comparable to or shorter than the period of the THz probe field. Such rapid modulation provides a unique opportunity to access and control EP physics within an ultrashort temporal window.
By varying the time delay $\Delta t$ between the optical pump and the THz probe pulses, the system can be selectively driven into either an EP or a non-EP regime.

This pump–probe delay thus acts as an effective ultrafast control knob, enabling dynamic switching between EP and non-EP states within a single excitation cycle. The ability to {dynamically} {parametrically} encircle exceptional points on picosecond timescales highlights the potential of optically pumped Ge as a platform for ultrafast, reconfigurable non-Hermitian photonic devices.

An experimental demonstration of this is shown in Fig.~\ref{fig:exp_fast}(a) where we plot the modulation depth at $F=259\mu\rm{J/cm}^2$ in co-polarization channel $-\Delta T/T$ with the delay time $\Delta t$ which is taken to be zero at the maxima. The big orange, red, and purple circular markers indicate $\Delta t=0.0,~0.5$ and $1.0$ ps over which the switching and encirclement are performed. Ultrafast switching through EP is shown in the inset, where the probe frequency is fixed at $\omega_{\rm{EP}}=0.6483$ THz. This value is lower than the simulated $\omega_{\rm{EP}}=0.7308$ THz because of polarizer effect explained in SM~\cite{SM} S4. The eigenstates merge at $\Delta t=0.5$ ps, marking the transition through EP. The corresponding transmission eigenvalue gaps become zero as shown in Fig.~\ref{fig:exp_fast}(b) and have the square root splitting, a hallmark of second-order EP as shown in Fig.~\ref{fig:exp_fast}(c). We also present a comparison of encirclement times from recent experiments employing THz OPTP setups with metamaterials (see Table 1 of SM~\cite{SM}~S7). The full encirclement path is shown in Fig.~\ref{fig:exp_fast}(d) on the 2D eigenvalue surfaces with $(\omega,\Delta t)$ parameter space. As one starts from the starting point and encircles the path shown with green dashed lines in the direction marked by arrows, one reaches the upper energy surface and doesn't return to the same eigenvalue, showing a clear verification of the topologically non-trivial nature of the second-order EP. Fig.~\ref{fig:exp_fast}(e) shows the variation of polarization eigenstates on the Poincaré sphere during the encirclement. The solid and dashed lines represent the eigenstates $\vert\psi_{\pm}\rangle$ respectively, varying with probe frequency for different delay times $\Delta t=0.0$ (orange) and $1.0$ ps (violet). As indicated by arrows, starting from $\vert\psi_+\rangle$, one reaches $\vert\psi_-\rangle$, while the $\vert\psi_-\rangle$ evolves into $\vert\psi_+\rangle$, marking the flipping or swapping of the eigenstates. Fig.~\ref{fig:exp_fast}(f) shows different polarization eigenstates, represented as polarization ellipses, as they evolve through the encirclement path around the EP.

To further quantify the non-orthogonality of transmission eigenmodes at EP, we calculate the Petermann factor defined as
\begin{equation}
K=\frac{\langle\psi^L\vert\psi^L\rangle\langle\psi^R\vert\psi^R\rangle}{\vert\langle\psi^L\vert\psi^R\rangle\vert^2},
\end{equation}
At EP, the eigenvectors become self-orthogonal, i.e, $\langle\psi^L\vert\psi^R\rangle\vert=0$,  leading to a divergence of $K$. In Fig.~\ref{fig:exp_fast}(g), the experimentally obtained value of $K$ is plotted across the parameter space, where the star-marked point exhibits a value of $K\approx 1365$ in the vicinity of the EP, signifying a pronounced non-Hermitian singularity. To further characterize the associated topological properties of this singularity, we analyze the eigenvalue gap $\Delta\lambda=\sqrt{S(\bm{R})}$, which exhibits non-analytic behavior and serves as a direct signature of the underlying topological braiding process, while remaining insensitive to global phase shifts~\cite{Kawabata2019}. In contrast, the squared gap,  $\Delta\lambda^2=S(\bm{R})$, is analytic in parameter space for a second-order exceptional point. Fig.~\ref{fig:exp_fast}(h) shows the variation of eigenvalue gap $\sqrt{S(\mathbf{R})}$ (as orange path) and its square $S(\mathbf{R})$ (blue path) as the parameters traverse one full cycle along the green rectangular loop shown in Fig.~\ref{fig:exp_fast}(g). Notably, the orange trajectory does not form a closed loop, whereas the blue trajectory does. This behavior results in accumulated topological phases of $180^\circ$ and $360^\circ$, respectively, as depicted in Fig.~\ref{fig:exp_fast}(i). These observations confirm the nontrivial topology of the transmission eigenvalues at EP, characterized by a half-integer (1/2) winding number.

It is worth noting that the modulation depth for the cross-polarization channel is optimum when $\vert T_{rl}\vert$ is minimized (See SM~\cite{SM} Fig.~S7. It is defined as $M=\vert\Delta T_{rl}\vert/T^{\rm{off}}_{rl}\vert\times 100$, where $\Delta T_{rl}=T^{\rm{off}}_{rl}-T^{\rm{on}}_{rl}$ is the difference in transmission with pump on and off. Our setup achieves a maximum modulation depth of $M=99\%$, detailed plots for the same are presented in SM~\cite{SM}~S7.

\section{Discussion}
In this work, we have achieved ultrafast active tuning and ultrafast parametric encirclement of an exceptional point (EP) in a 2D metasurface composed of Ge-layered split-ring resonators. By employing the optical pump–terahertz probe (OPTP) technique, we systematically encircle the EP through controlled variation of the pump fluence and probe frequency. Through optimized active tuning, we realize an absolute eigentransmission gap on the order of $\approx 10^{-2}$corresponding to a significantly enhanced Petermann factor of $K\approx 1365$ experimentally. This is accompanied by a high cross-polarization modulation depth of up to $99\%$. Furthermore, by {dynamically} {parametrically} sweeping the pump–probe delay, we demonstrate ultrafast parameter encirclement, completing a full loop within $1$ ps {delay window}. To our knowledge, this is the fastest EP encirclement and switching time (SM~\cite{SM}~S7) and modulation depth achieved to date in experiments. This ultrafast dynamics directly arises from the carrier dynamics in the Ge layer, which has a very short lifetime. Our results align well with theoretical predictions from both the dipole model and germanium carrier dynamics models. We also demonstrate that accounting for imperfect polarizers in simulation modeling can cause slight shifts in the exceptional point (EP) position within parameter space and on the Poincaré sphere (SM ~\cite{SM}~S4). These results have strong implications for the development of ultrafast and highly robust topological state manipulation~\cite{Said2022} and information processing in photonic and quantum devices~\cite{Li2024} via polarization modulators and switches in the THz regime. Also, the fine active control enables us to reach very close to EP. This opens a pathway for building ultrasensitive sensors for probing the refractive index modulation, chiral biomolecule detection, and precise identification of polarization states. Further applications can be developed in beam deflection~\cite{Chen2022,Zhao2022,Wang2025}, THz holography~\cite{Heimbeck2020,Song2021,Yang2024,Wu2024}, and imaging for early cancer detection~\cite{Sadeghi2023}.

\section{Methods}

\textbf{\textit{Sample Fabrication}}:
The device is fabricated on a 1 mm-thick quartz substrate, where orthogonally oriented split-ring resonators (SRRs) are patterned using direct-write optical lithography (Microwriter, $\lambda$ = 405 nm). A Z-cut single-crystal quartz wafer was selected as the substrate. To enable a clean lift-off process, a bilayer resist stack consisting of LOR and a positive photoresist (PPR) was employed. Following patterning, a Ti/Au bilayer (15/200 nm) was deposited by sputtering, with Ti serving as the adhesion layer. After lift-off, a 100 nm-thick germanium (Ge) film was deposited on top of the resonators by electron-beam evaporation to ensure uniform coverage, acting as the optically pumped active medium.

\textbf{\textit{Comsol simulations}}: The design of metasurface is optimized using FEM simulations. We used transition boundary conditions and floquet periodic boundary conditions to obtain the complex transmission spectra. The geometrical parameters of SRRs are optimized to enable maximum asymmetric transmission, minimum eigenenergy gap, for potential use as a polarization filter.

\textbf{\textit{THz OPTP measurements}}:
The terahertz (THz) pulses are generated using optical rectification in a Zinc Telluride (ZnTe) crystal. A femtosecond pulsed laser with a repetition rate of 1 kHz and a pulse duration of 35 femtoseconds (fs) is used as the optical source. The laser beam is initially split by a beamsplitter (BS1, R70:T30) into high-intensity beam and a low-intensity beam. The high-intensity beam is used as the optical pump to photoexcite carriers in the Ge sample. The delay time between pump and probe beam is controlled using a mechanical delay stage, enabling precise adjustment of the pump–probe delay. The low-intensity beam is further split by a second beamsplitter (BS2, R90:T10). The reflected portion is directed onto the ZnTe crystal to generate the THz pulse. The generated THz radiation is collimated and focused onto the sample using a pair of off-axis parabolic mirrors. The transmitted THz pulse carries the information about the photoinduced carrier dynamics in the sample. The transmitted portion of BS2 served as the optical probe beam. Its arrival time at the detection crystal relative to the THz pulse is controlled by a separate probe delay stage. The probe beam and the transmitted THz pulse are spatially and temporally overlapped inside a second ZnTe crystal for electro-optic sampling (EOS). When the THz electric field co-propagates with the probe pulse through the ZnTe detection crystal, it induces a transient birefringence via Pockels effect. This birefringence causes a polarization rotation in the probe pulse proportional to the instantaneous THz electric field. A quarter-wave plate followed by a Wollaston prism separates the probe beam into orthogonal s- and p-polarization components. The intensities of these two polarization components are detected using a balanced photodiode detector, which measures the differential signal between the s and p components.
To obtain all co- and cross-polarization components of the transmission matrix, polarization-resolved measurements are performed using wire-grid polarizers (WGPs). WGP2 is fixed at 45° so that the orthogonal component of electric field can be projected onto an identical output polarization. This configuration ensures that the ZnTe detector responds equivalently to all incident polarization states, providing consistent detection sensitivity across measurements. The remaining polarizer, WGP1, together with the sample orientation, is rotated between 0° and 90° to measure the individual transmission matrix elements. Specifically, the $T_{xx}$ component is obtained with both the sample and WGP1 aligned along 0°, while $T_{yy}$ is measured with the sample rotated to 90° keeping WGP1 fixed at 0°. For the cross-polarized components, $T_{xy}$ is measured with the sample at 0° and WGP1 at 90°, whereas $T_{yx}$ is obtained with the sample at 90° and WGP1 at 90°. The entire experimental setup is enclosed in a nitrogen-purged environment to minimize absorption of THz radiation by atmospheric water vapor and to ensure stable measurement conditions. A schematic of the detailed experimental setup is provided in the SM~\cite{SM} Sec.~S1.

\section{Acknowledgment}

P.S. acknowledges the support by the Prime Minister Research Fellowship (PMRF), Govt. of India. We acknowledge funding support from the National Quantum Mission, an initiative of the Department of Science and Technology, Government of India. A.K. acknowledges the support from ANRF via grant number CRG/2022/001170.

\bibliography{references}

\end{document}